\documentclass[preprint,showpacs,superscriptaddress,preprintnumbers,longbibliography]{revtex4-1}
\usepackage{amssymb}
\usepackage{amsmath}    % need for subequations
\usepackage{graphicx}   % need for figures
\usepackage{verbatim}   % useful for program listings
\usepackage{color}      % use if color is used in text
\usepackage{ulem}
\usepackage{subfigure}  % use for side-by-side figures
\usepackage[english]{babel}
\usepackage{hyperref}   % use for hypertext links, including those to external documents and URLs
\usepackage{float}
\usepackage{placeins}

\begin{document}
\title{Emergence of the strong tunable linear Rashba spin-orbit coupling of two-dimensional hole gases in semiconductor quantum wells}
\author{Jia-Xin Xiong}
\affiliation{State Key Laboratory of Superlattices and Microstructures, Institute of Semiconductors,Chinese Academy of Sciences, Beijing 100083, China}
\affiliation{Center of Materials Science and Optoelectronics Engineering, University of Chinese Academy of Sciences, Beijing 100049, China}
\author{Shan Guan}
\email{shan\_guan@semi.ac.cn}
\affiliation{State Key Laboratory of Superlattices and Microstructures, Institute of Semiconductors,Chinese Academy of Sciences, Beijing 100083, China}
\author{Jun-Wei Luo}
\email{jwluo@semi.ac.cn}
\affiliation{State Key Laboratory of Superlattices and Microstructures, Institute of Semiconductors,Chinese Academy of Sciences, Beijing 100083, China}
\affiliation{Center of Materials Science and Optoelectronics Engineering, University of Chinese Academy of Sciences, Beijing 100049, China}
\affiliation{Beijing Academy of Quantum Information Sciences, Beijing 100193, China}
\author{Shu-Shen Li}
\affiliation{State Key Laboratory of Superlattices and Microstructures, Institute of Semiconductors,Chinese Academy of Sciences, Beijing 100083, China}
\affiliation{Center of Materials Science and Optoelectronics Engineering, University of Chinese Academy of Sciences, Beijing 100049, China}

\begin{abstract}
{\textbf{Two-dimensional hole gases in semiconductor quantum wells are promising platforms for spintronics and quantum computation but suffer from the lack of the $\bf{k}$-linear term in the Rashba spin-orbit coupling (SOC), which is essential for spin manipulations without magnetism and commonly believed to be a $\bf{k}$-cubic term as the lowest order. Here, contrary to conventional wisdom,  we uncover a strong and tunable $\bf{k}$-linear Rashba SOC in  two-dimensional hole gases (2DHG) of semiconductor quantum wells by performing atomistic pseudopotential calculations combined with an effective Hamiltonian for a model system of Ge/Si quantum wells. Its maximal strength exceeds 120 meV{\AA}, comparable to the highest values reported in narrow bandgap III-V semiconductor 2D electron gases, which suffers from short spin lifetime due to the presence of nuclear spin.  We also illustrate that this emergent $\bf{k}$-linear Rashba SOC is a first-order direct Rashba effect, originating from a combination of heavy-hole-light-hole mixing and direct dipolar intersubband coupling to the external electric field. These findings confirm Ge-based 2DHG to be an excellent platform towards large-scale quantum computation.
}}\end{abstract}

\maketitle
Spin-orbit coupling (SOC) entangles the spin and orbital degrees of freedom and has inspired a vast number of predictions, discoveries and innovative concepts, including spin transistors, spin-orbit qubits, spin Hall effect, quantum spin Hall effect, topological insulators, and Majorana fermions\cite{Datta1990, Schliemann2003, Maurand2016, Watzinger2018, Sinova2004, DiXiao2010, Bernevig2005, Bernevig2006, Awschalom2013, Sarma2010, Mourik2012, Manchon2015}. The exploration, understanding, and control of SOC have become intensive research subjects across many different disciplines in condensed matter physics. Very recently, the strong Rashba SOC of holes in the platform of Ge quantum wells (QWs) has been demonstrated to provide an efficient driving manner for rapid qubit control~\cite{Hendrickx2020}. The electric-field tunability of the Rashba SOC further ensures the independent control of multiple qubits, which obviates the need for microscopic elements close to each qubit~\cite{Hendrickx2020}. In contrast, the absence of strong SOC in Si demands the inclusion of complicated components in the proximity of each qubit to control the qubit, making the scalability of Si qubits being a key challenge\cite{Hendrickx2020}, despite Si qubits so far have been considered as the most promising platform for large-scale quantum computation\cite{Awschalom2013, Hendrickx2020}. Therefore, in the combination of holes being free from the challenge of valley degeneracy for electrons in Si platform as well as Ge possessing the highest hole mobility among all known semiconductors, reaching a hole mobility over 1.5 million cm$^2$/(V$\cdot$s) at 3 K in strained Ge QWs\cite{Failla2016}, strong SOC of holes renders Ge QWs as the excellent platform for large-scale quantum computation.

\begin{figure*}[!hbp]
\centering
\includegraphics[width=\linewidth]{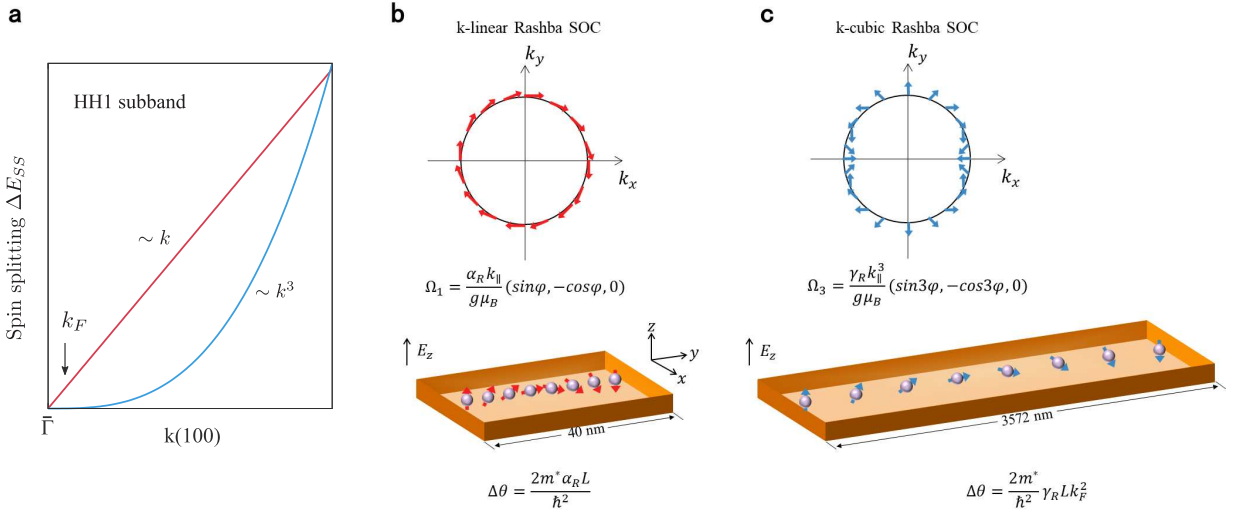}
\caption{Schematically illustration of the compelling different features between $\bf{k}$-linear (marked in red) and $\bf{k}$-cubic (marked in blue) Rashba SOC. $\bf{a}$, Rashba spin splitting of the ground heavy hole (HH) subband  around $\bar{\Gamma}$ point in semiconductor quantum wells. It is commonly believed that the Rashba spin splitting should be $\bf{k}$-cubic, whereas we unravel that it should be $\bf{k}$-linear as the lowest order. {\bf{b}}, The effective magnetic field texture of $\bf{k}$-cubic Rashba SOC in the momentum space, and the corresponding  spin precession $\Delta \theta$ induced by the  $\bf{k}$-cubic Rashba SOC in the real space. The magnetic field of the $\bf{k}$-linear Rashba SOC must orient perpendicularly to the momentum. {\bf{c}}, The effective magnetic field texture of $\bf{k}$-linear Rashba SOC in the momentum space, and the corresponding  spin precession $\Delta \theta$ induced by the  $\bf{k}$-linear Rashba SOC in the real space.  Here $\alpha_R$ ($\gamma_R$) is the $\bf{k}$-linear ($\bf{k}$-cubic) Rashba parameter, $\Omega_1$ ($\Omega_3$) the $\bf{k}$-linear ($\bf{k}$-cubic) effective magnetic field, $k_F$ the Fermi wave vector, $L$ the spin precession length of hole spin transistors, $m^*$ the hole effective mass (we take $0.3m_0$ for Ge, where $m_0$ is the mass of bare electron). The expressions of the spin precession angle based on the $\bf{k}$-linear and $\bf{k}$-cubic Rashba SOC are given in Ref.\cite{Datta1990} and Ref.\cite{Marco2004}, respectively.  For the $\bf{k}$-cubic SOC with a Rashba parameter $\gamma_R=2.26\times 10^5$ meV\AA$^3$ (value adopted from Ref.\cite{Moriya2014} for Ge) and the Fermi wave vector is $0.002\times 2\pi/a$ ($a=5.65$ {\AA} is the lattice constant of Ge), the spin precession length will be 3.6 $\mu m$. However, for the $\bf{k}$-linear SOC with a Rashba parameter $\alpha_R=100$ meV{\AA} (value within the range of our SEPM results), the spin precession length will be 40 nm, which is two orders of magnitude smaller than the length of the $\bf{k}$-cubic Rashba SOC.  }
\label{fig1}
\end{figure*}

However, the Rashba SOC of the ground hole subband in semiconductor QWs, including Ge QWs, is commonly believed to be $\bf{k}$-cubic rather than $\bf{k}$-linear as the lowest order\cite{Winkler2003,Bulaev2005,EDSR2007}. This is in sharp contrast to the electron counterpart, in which a variety of potential applications has been proposed based on the $\bf{k}$-linear term. We schematically illustrate the  different features between the $\bf{k}$-cubic and $\bf{k}$-linear Rashba SOC in Figure \ref{fig1}. Compared to the $\bf{k}$-linear spin splitting, the $\bf{k}$-cubic spin splitting is negligible at the Fermi wave vector $k_F$, which is around $\bar{\Gamma}$ point due to the light doping. Subsequently, from the perspective of spin splitting,  the $\bf{k}$-linear or  $\bf{k}$-cubic Rashba SOC is  a ``yes-or-no" question rather than a ``large-or-small" question. Secondly, as indicated in Figure \ref{fig1}b, \ref{fig1}c, the effective magnetic field texture in the momentum space is quite different. Due to different rotation rates, only the effective magnetic field of the $\bf{k}$-linear Rashba SOC is ``locked" perpendicularly to the momentum, providing more precise control over the spin direction. Thirdly, in real space, as shown in Figure \ref{fig1}b, \ref{fig1}c, the $\bf{k}$-linear Rashba SOC induces a much faster spin precession than the $\bf{k}$-cubic Rashba SOC, leading to two orders of magnitude smaller device length. This advantage caters to the needs of short-size hole spin transistors.

Because of these compelling differences between $\bf{k}$-linear and $\bf{k}$-cubic terms, the absence of strong $\bf{k}$-linear Rashba SOC excludes holes in QWs from many potential applications. Luo et al.\cite{Luo2011} and Kloeffel et al.\cite{Kloeffel2011} independently found, in one-dimensional (1D) quantum wires, the emergence of a strong $\bf{k}$-linear hole Rashba SOC, originating from a direct dipolar coupling between heavy-hole (HH) and light-hole (LH) subbands by an external electric field\cite{Kloeffel2011,Kloeffel2018a,Luo2017}. Such a giant $\bf{k}$-linear Rashba effect, called the direct Rashba effect, is a first-order effect and much stronger than the conventional third-order Rashba effect. The strength of the direct Rashba SOC scales with the HH-LH coupling at zone center (${\bf \bar k}=0$) and is thus supposed to vanish in two-dimensional (2D) QWs since the HH-LH coupling is commonly considered to be forbidden by symmetry\cite{Winkler2003}. However, the zone-center HH-LH coupling was already found experimentally and theoretically in QWs\cite{Luo2010,Ivchenko1996,Krebs1996,Toropov2000,Durnev2014,luo_supercoupling_2015,Sherman1988}. One may naturally expect the existence of a strong $\bf{k}$-linear hole Rashba SOC in QWs. Such expectation is relevant to the current understanding of 2D hole spin physics, such as spin-Hall conductivity\cite{Schliemann2004,Sugimoto2006,Wong2010,Bi2013, Bernevig2005}, spin-galvanic effect\cite{Ganichev2002}, hole spin helix~\cite{Sacksteder2014} and current-induced spin polarization~\cite{Liu2008a}, since all these effects have been investigated based on the assumption of the $\bf{k}$-cubic Rashba SOC\cite{Marcellina2017}.

In this article, we study the Rashba SOC-induced spin-splitting in Ge/Si QWs using the atomistic semi-empirical pseudopotential method (SEPM)\cite{Wang1994,Wang1995,Wang1999} in combination with the theoretical analysis based on the Luttinger-Kohn Hamiltonian. We indeed find finite $\bf{k}$-linear Rashba SOC even in [001]-oriented Ge/Si QWs, which is confirmed to be a direct Rashba SOC based on an effective Hamiltonian considering a finite HH-LH mixing. Moreover, a much stronger $\bf{k}$-linear Rashba SOC is observed in [110]-oriented Ge/Si QWs. The predicted Rashba parameter $\alpha_R$ can be tuned as large as 120 meV{\AA}, which is among the largest values measured in 2D electron systems\cite{Manchon2015}. This electric-field tunability of the 2D $\bf{k}$-linear Rashba SOC, over the broad range of two orders of magnitude, demonstrates its predominance over 2D $\bf{k}$-linear Dresselhaus SOC\cite{Luo2010}, providing a promising platform for hole spin transistors.

\vspace*{5mm}
\noindent {\bf \large Results}

\begin{figure}[!hbp]
\centering
\includegraphics[width=0.8\linewidth]{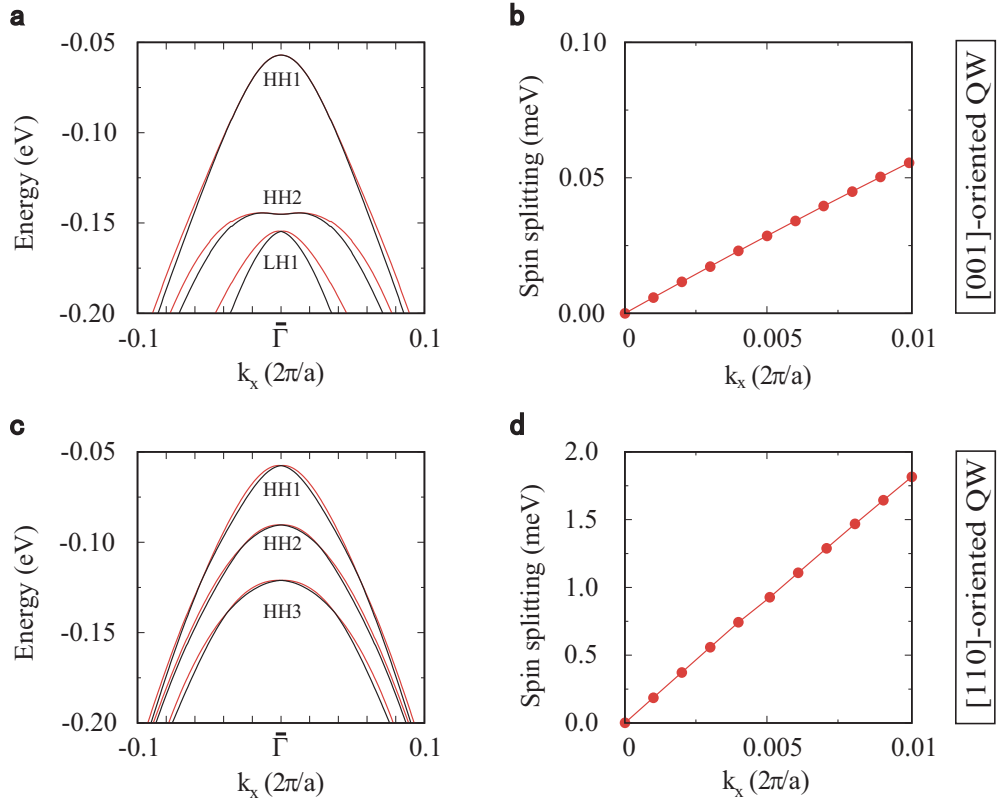}
\caption{Calculated energy dispersion of valence subbands and spin splitting of HH1 for $\bf{a}$, $\bf{b}$ [001]-oriented (Ge)$_{40}$/(Si)$_{20}$ QW and $\bf{c}$, $\bf{d}$ [110]-oriented (Ge$)_{40}$/(Si)$_{20}$ QW, respectively, under an electric field $E_z=100$ kV/cm perpendicular to the interface. Here, the thickness (subscripts) units in monolayer (ML).  The $x$-direction in [001]- and [110]-oriented QWs is along with the crystalline [100] and [001] directions, respectively. The labels HH1, HH2, HH3, and LH1 indicate the valence subbands derived mainly from either bulk HH or LH bands.}
\label{fig2}
\end{figure}

\noindent
\textbf{The k-linear Rashba spin splitting.} Figure \ref{fig2}a, \ref{fig2}c show the atomistic SEPM calculated valence band structure of [001]- and [110]-oriented Ge/Si QWs upon application of an electric field of 100 kV/cm normal to the QWs. The states of the conduction subbands are confined in Si layers while states of the valence subbands are confined in Ge layers (not shown) due to the bulk Ge valence band maximum (VBM) being 0.5 eV higher than the bulk Si\cite{Winkler2003}. We find that the doubly spin-degenerate subbands split away from the $\bar{\Gamma}$ point giving rise to spin-splitting $\Delta E_{ss}(k_{\|})$, which is shown in Figure \ref{fig2}b, \ref{fig2}d for the ground state subband derived from the bulk HH band (HH1). $\Delta E_{ss}(k_{\|})$ exhibits a nice linear scale against $k_x$. Due to the existence of an inversion center, the bulk inversion asymmetry-induced Dresselhaus spin splitting\cite{Dresselhaus1955} is absent in Ge/Si QWs. Hence, the obtained spin splitting is completely induced by the Rashba effect\cite{Rashba1984}. Interestingly, we find that $\alpha_{R}$ is anisotropic in [110]-oriented QWs but isotropic in [001]-oriented QWs. Such anisotropic $\bf{k}$-linear Rashba SOC in the [110]-oriented QWs is against common sense that $\bf{k}$-linear Rashba SOC is always isotropic in 2D systems. This anisotropy is due to the breaking of the axial symmetry and will be explained below. To deduce the linear ($\alpha_{R}$) and cubic ($\gamma_{R}$) Rashba parameters, we fit spin-splitting $\Delta E_{ss}(k_{\|})$ of the HH1 subband to the equation $\Delta E_{ss}(k_{\|}) = 2\alpha_{R}k_{\|} + \gamma_{R}k^3_{\|}$, obtaining $\alpha_{R}=3,\ 82$ meV{\AA} for [001]- and [110]-oriented (Ge)$_{40}$/(Si)$_{20}$ QWs (under an external field of 100 kV/cm), respectively. The value of [110]-oriented QWs is comparable to that of direct hole Rashba SOC predicted in 1D quantum wires\cite{Luo2017}.

\vspace*{2mm}

\begin{figure}[!hbp]
\centering
\includegraphics[width=0.8\linewidth]{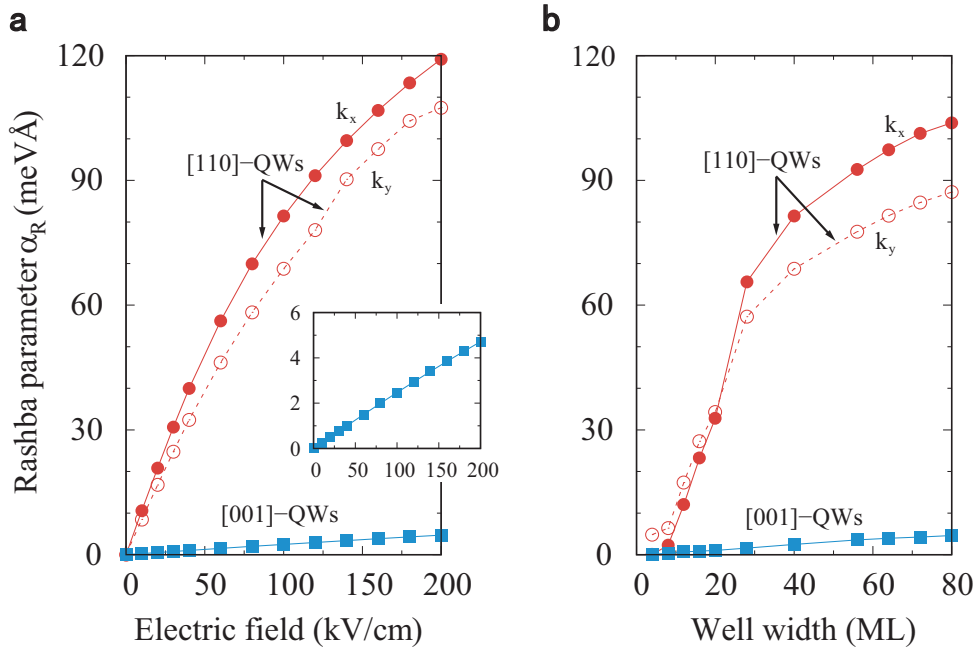}
\caption{Calculated $\bf{k}$-linear hole Rashba parameters $\alpha_{R}$ in Ge/Si QWs as a function of $\bf{a}$ electric field strength with 40 ML well thickness, and $\bf{b}$ well thickness under an electric field of 100 kV/cm, respectively. The Si thickness is 20 ML. For [110]-oriented QWs, the $x$- and $y$-direction is along with the crystalline [001] and [1$\bar{1}$0] direction, respectively. The inset to  $\bf{a}$ shows a linear dependence in [001]-oriented QWs, reflecting the negligible QCSE in comparison to QCE. Note that a larger electric field strength than 200 kV/cm is not experimentally available, hence the results are not shown.}
\label{fig3}
\end{figure}

We examine the field- and size-dependences of the $\bf{k}$-linear Rashba parameter $\alpha_R$ for both [001]- and [110]-oriented QWs. Figure \ref{fig3}a shows that $\alpha_{R}$ scales linearly as $E_z$ for the [001]-oriented (Ge)$_{40}$/(Si)$_{20}$ QW, but scales sublinearly for the [110]-oriented (Ge)$_{40}$/(Si)$_{20}$ QW, in which the difference of $\alpha_{R}$ along $k_x$ and $k_y$ directions gets bigger with increasing $E_z$. Under a fixed electric field $E_z=100$ kV/cm with varying well thickness, as shown in Figure \ref{fig3}b, $\alpha_{R}$ increases linearly against the well thickness for [001]-oriented QWs. Whereas, for [110]-oriented QWs, $\alpha_{R}$ increases linearly in a much larger rate in well thickness when $L<20$ ML, and then grows slowly towards saturation. The difference of $\alpha_{R}$ along $k_x$ and $k_y$ directions is negligible when $L<20$ ML but raises quickly with further increasing $L$. Such field- and size-dependences of $\alpha_{R}$ in [110]-oriented QWs are similar to the case of 1D quantum wires~\cite{Luo2017}, indicating spin-splitting arising from the first-order direct Rashba SOC rather than conventional third-order Rashba SOC. Interestingly, $\alpha_R^{[110]}$ is one order of magnitude larger than $\alpha_R^{[001]}$, and is strongly tunable by external field to exceed 120 meV{\AA}. This strong and tunable Rashba SOC in  [110]-oriented QWs is a compelling property for hole spin manipulation.

\vspace*{2mm}

\noindent
\textbf{The origin of k-linear Rashba SOC.} We turn to unravel the origin of the emergence of $\bf{k}$-linear hole Rashba SOC and illustrate the decisive role of the HH-LH mixing through the envelope function approximation based on the Luttinger-Kohn (LK) Hamiltonian. The lowest-energy subband spectrum is governed by the effective Hamiltonian projected into the subspace spanned by the four states $|\textrm{HH1}_{\pm}\rangle$ and $|\textrm{HH2}_{\pm}\rangle$ of the two topmost HH-like subbands (indicated in the  atomistic SEPM band structure by HH1 and HH2) at $\bar{\Gamma}$-point. The four basis states are constructed by including the HH-LH mixing: $|\textrm{HH1}_{\pm}\rangle=a_1\phi_1(z)|\frac{3}{2},\pm\frac{3}{2}\rangle+b_1\phi_1(z)|\frac{3}{2},\mp\frac{1}{2}\rangle$ and $|\textrm{HH2}_\pm\rangle=a_2\phi_2(z)|\frac{3}{2},\pm\frac{3}{2}\rangle+b_2\phi_2(z)|\frac{3}{2},\mp\frac{1}{2}\rangle$, where $a_{1}$ and $b_{1}$ ($a_{2}$ and $b_{2}$) are real coefficients of bulk HH and LH Bloch functions in QW HH1 (HH2) states, respectively, and  envelope functions $\phi_n(z)=\sqrt{\frac{2}{L}}sin[\frac{n\pi(z+L/2)}{L}]$, n$\in$\{1,2,$\cdots$\}, by assuming an infinite confinement potential.

In the [001]-oriented QWs with $D_{2d}$ symmetry, the HH-LH mixing at $\bar{\bf{k}}=0$ originates from symmetry reduction caused by the local symmetry of the $C_{2v}$ interface\cite{Ivchenko1996, luo_supercoupling_2015}, otherwise is absent (i.e., $b_{1,2}=0$). Starting from the [001]-oriented 3D LK Hamiltonian\cite{Luttinger1956,Kloeffel2011}, we obtain the $4\times 4$ effective Hamiltonian of 2D  [001]-oriented QWs in the absence of the external electric field,
\begin{equation}\label{Rashba1}
H_{\textrm{eff}}^{[001]}=A_++A_-\tau_z+\gamma_3C_0\tau_x(k_x\sigma_y-k_y\sigma_x),
\end{equation}
where $\tau$ and $\sigma$ are the Pauli matrices describing the orbital part and the spin part of eigenstates, respectively, and $\gamma_3$ the LK parameter. $A_{\pm}=\hbar^2k_x^2(m_{1x}^{-1}\pm m_{2x}^{-1})/4+\hbar^2k_y^2(m_{1y}^{-1}\pm m_{2y}^{-1})/4\pm \Delta_0/2$, where $m_{1x}$ and $m_{1y}$ ($m_{2x}$ and $m_{2y}$) are  effective masses along $x$- and $y$-direction, respectively, for HH1 (HH2) subband and $\Delta_0$ the energy separation between HH1 and HH2 states arising from the space confinement effect (SCE). The off-diagonal term $\langle \textrm{HH1}_{\pm}|H_{\textrm{eff}}^{[001]}|\textrm{HH2}_{\mp} \rangle=\mp i\gamma_3C_0(k_x\mp ik_y)$, where the coupling parameter $C_0=(a_1b_2-a_2b_1)\frac{8\sqrt{3}\hbar^2}{3m_0L}$ reflects the strength of HH-LH mixing.

Upon application of an external electric field $E_z$ to [001]-oriented QWs, $E_z$  will couple directly to the spins owing to HH-LH mixed QW states, yielding a direct dipolar coupling term $\langle \textrm{HH1}_{\pm}|(-eE_zz)|\textrm{HH2}_{\pm} \rangle = eE_zU_0$, where the coupling constant $U_0=(a_1a_2+b_1b_2)\frac{16L}{9\pi^2}$ is also related to the HH-LH mixing. Using quasi-degenerate perturbation theory \cite{Schrieffer1966}, we finally obtain the first-order $2\times2$ effective Rashba SOC Hamiltonian for the HH1 subband: $H_{soc}^{[001]}=\alpha_R^{[001]}(k_x\sigma_y-k_y\sigma_x)$ (see supplemental material for details), where the $\bf{k}$-linear Rashba parameter reads
\begin{equation}\label{Rashba2}
\alpha_R^{[001]}=\frac{2e\gamma_3C_0U_0E_z}{\sqrt{\Delta_0^2+4e^2U_0^2E_z^2}}.
\end{equation}
The denominator term is the energy separation  $\Delta E_{1,2}$ between HH1 and HH2 induced by SCE ($\Delta_0$) and quantum-confined Stark effect (QCSE) ($2eU_0E_z$). It is straightforward to learn that $\alpha_R^{[001]}$ scales linearly with $E_z$ when $2eU_0E_z\ll\Delta_0$, in excellent agreement with our atomistic SEPM results shown in Figure \ref{fig3}a. The  ${\bf k}$-linear Rashba SOC originates from  a combination of the HH-LH mixing and the direct dipolar coupling to the external electric field, with $\alpha_R^{[001]}$ having the same formula [Eq. \ref{Rashba2}] as that of the direct Rashba SOC in 1D nanowires~\cite{Kloeffel2011}. Hence, the ${\bf k}$-linear Rashba SOC uncovered in [001]-oriented QWs is a 2D direct Rashba effect.

In the [110]-oriented QWs, besides the interface-induced HH-LH mixing, the breaking of the axial symmetry causes an intrinsic HH-LH mixing at $\bf{\bar k}=0$\cite{Winkler2003} with its magnitude proportional to $(\gamma_3-\gamma_2)\times \hat{k}_z^2$, where $\hat{k}_z^2\sim (\pi/L)^2$\cite{Kloeffel2018a}. This intrinsic HH-LH mixing leads to an enhanced direct Rashba effect in [110]-oriented QWs in comparison to [001]-oriented QWs as we have observed in our atomistic SEPM results (Figure \ref{fig3}). We perform the same procedure as done in [001]-oriented QWs and obtain an in-plane anisotropic linear Rashba SOC (see supplemental material for details),
\begin{equation}\label{Rashba3}
\alpha_{R}^{[110]}(k_x)=\frac{2e\gamma_3C_0U_0E_z}{\sqrt{{\Delta_0}^2+4e^2{U_0}^2E_z^2}},
\end{equation}
and
\begin{equation}\label{Rashba4}
\alpha_{R}^{[110]}(k_y)=\frac{2e\gamma_2C_0U_0E_z}{\sqrt{{\Delta_0}^2+4e^2{U_0}^2E_z^2}}.
\end{equation}
$\alpha_{R}^{[110]}(k_x)/\alpha_{R}^{[110]}(k_y)\approx \gamma_3/\gamma_2$ (the Luttinger parameters for Ge are $\gamma_3$=5.69 and $\gamma_2$=4.24\cite{Marcellina2017}, hence $\alpha_R^{[110]}(k_x)/\alpha_R^{[110]}(k_y)\approx 1.34$), which explains the atomistic SEPM results shown in Figure \ref{fig3}.

\begin{figure}[!hbp]
\centering
\includegraphics[width=0.8\linewidth]{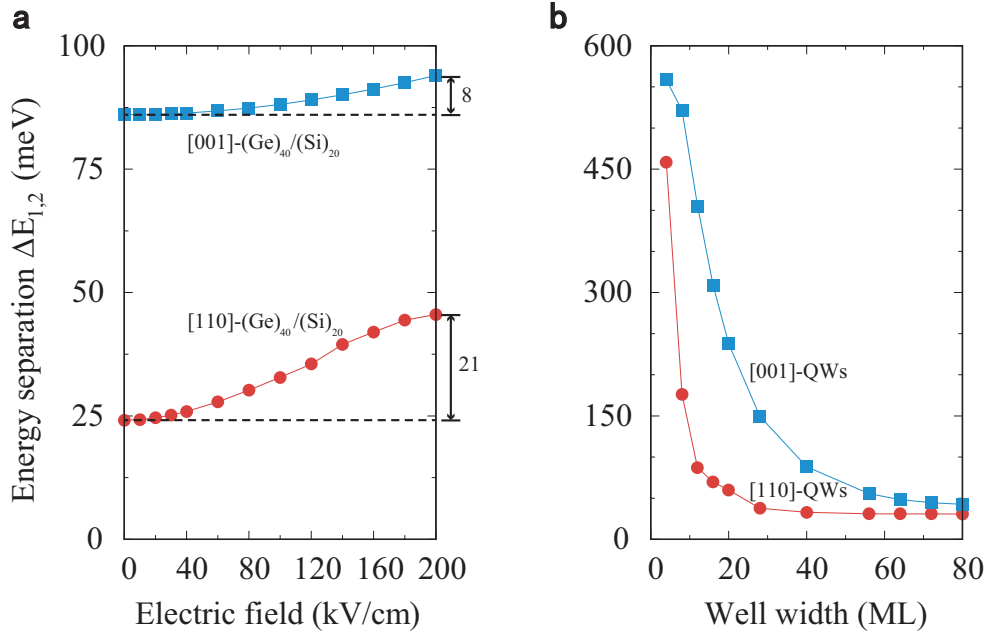}
\caption{Energy separation $\Delta E_{1,2}=\sqrt{\Delta_0^2+4e^2U_0^2E_z^2}$ between HH1 and HH2 states at $\Gamma$ point in [001]- and [110]-oriented Ge/Si QWs as a function of $\bf{a}$ electric field strength for a fixed QW thickness and $\bf{b}$ well thickness with an applied electric field of 100 kV/cm, respectively.}
\label{fig4}
\end{figure}
\vspace*{2mm}
\noindent
\textbf{The reason of much larger Rashba SOC in [110]-oriented QWs.} To understand the observed different behaviors of [001]- and [110]-oriented QWs in the field- and size-dependence of Rashba SOC strength ($\alpha_R$ as shown in Figure \ref{fig3}), we examine the energy separation $\Delta E_{1,2}$ which is the denominator term in the expression of Rashba parameters (Eq.~\ref{Rashba2}-\ref{Rashba4}). Figure \ref{fig4}a shows $\Delta E_{1,2}$ as a function of the applied electric field $E_z$ for a fixed thickness $L=40$ ML, in which $\Delta E_{1,2}$  is solely induced by SCE at $E_z=0$. With increasing $E_z$, $\Delta E_{1,2}$ grows in substantially different rates for [001]- and [110]-oriented QWs. A much larger rate in [110]-oriented QWs implies a stronger QCSE, illustrating that the HH-LH mixing and thus the direct Rashba SOC could have a strong response to the external electric field. At $E_z$=200 kV/cm, we find that the QCSE contributes 50\% to $\Delta E_{1,2}$ in [110]-oriented QWs, but less than 10\% in [001]-oriented QWs. The significantly enhanced contribution of QCSE ($2eU_0E_z$) to $\Delta E_{1,2}$ causes a sub-linear field-dependence of $\alpha_R$ observed in [110]-oriented QWs [Eq. \ref{Rashba3} and Figure \ref{fig3}a]. Whereas, in [001]-oriented QWs, the weak field-dependence of $\Delta E_{1,2}$, implying a much weaker QCSE, leads to a linear scale of $\alpha_R$ as applied field $E_z$ [Eq. \ref{Rashba2} and Figure \ref{fig3}a]. Figure \ref{fig4}b displays $\Delta E_{1,2}$ against the well thickness $L$ under a fixed $E_z$. We see that, with increasing $L$, $\Delta E_{1,2}$ drops rapidly for $L\leqslant40$ ML and then decreases slowly towards the bulk zero value owing to the reduced SCE. Interestingly, in addition to the much stronger QCSE, [110]-oriented QWs have a 3-4 times smaller $\Delta E_{1,2}$ than [001]-oriented QWs. In combination with the enhanced HH-LH mixing originated from the breaking of the axial symmetry, [110]-oriented QWs perform an order stronger $\bf{k}$-linear Rashba SOC than [001]-oriented QWs.

\vspace*{5mm}
\noindent {\bf Potential applications of the k-linear Rashba effect.}
The $\bf{k}$-linear Rashba effect, as summarized in Table~\ref{table1}, plays a significant role in 2D electron systems. However, the absence of this $\bf{k}$-linear Rashba effect in 2D hole systems rules out its potential applications. For example, the interplay between Rashba and Dresselhaus effect with equal $\bf{k}$-linear parameters, i.e., $\alpha_R=\beta_D$, guarantees the spin precession robust against spin scatterings, leading to a long spin lifetime\cite{N.S.Averkiev1999,Schliemann2003}. This property, protected by SU(2) symmetry~\cite{J.D.Koralek2009}, is crucial to devices such as spin transistors\cite{Schliemann2003}. The past point of view that only the $\bf{k}$-cubic Rashba effect exists in 2D hole systems overlooks this $\bf{k}$-linear counterpart, thus misleading us ruling out the 2D hole systems for high-quality spin transport. Our discovery of the $\bf{k}$-linear Rashba spin splitting fills this void. The effects related to the $\bf{k}$-linear Rashba spin splitting where electrons have in 2D systems, become implementable for holes at present.

\begin{table*}[ht]
\caption{Physical effects related to spin-orbit interaction and their applications~\cite{Manchon2015}. Here, $\alpha_R$ and $\gamma_R$ denote $\bf{k}$-linear and k-cubic Rashba parameters, respectively; $\beta_D$ represents the $\bf{k}$-linear Dresselhaus parameter; $e$ is the elementary charge, $\hbar$ the reduced Plank constant, $m^*$ the effective mass; $\vec{\sigma}$ is the Pauli matrix, $\vec{p}$ the momentum.}
\newcommand{\tabincell}[2]{
        \begin{tabular}{@{}#1@{}}#2\end{tabular}
}
\begin{tabular}  {p{2.6cm} p{5.8cm} p{3.9cm} p{3.7cm}}
  \hline\hline
     Effect & Physical quantity & Formula  & Application \\ \hline
  Spin relaxation suppression\cite{N.S.Averkiev1999,Schliemann2003} & \tabincell{l}{$\bf{k}$-linear Rashba parameter $\alpha_R$,\\ Dresselhaus parameter $\beta_D$} & $\alpha_R=\beta_D$  & Persistent spin helix states, spin transistors \\ \hline
  Spin Hall effect (SHE) \cite{Sinova2004, DiXiao2010, Bernevig2005} & \tabincell{l}{Spin Hall conductivity $\sigma_H$,\\ Berry phase $\phi_B$,\\ Berry curvature $\vec{\Omega}_p$} & \tabincell{l}{$\sigma_H=\frac{e\phi_B}{8\pi^2}$,\\ $\phi_B=\oint_{S}\vec{\Omega}_p\cdot d\vec{S}$,\\ $\vec{\Omega}_p\propto\alpha_R\nabla_p\times(\vec{\sigma}\times\vec{p})$} & All-semiconductor spin Hall effect transistor, spin AND logic function \\ \hline
  Spin-galvanic effect\cite{J.Wunderlich2009,Ivchenko1978,Ganichev2008} & \tabincell{l}{Induced charge current density $\vec{j}_c$,\\ non-equilibrium spin density $\vec{S}$} & \tabincell{l}{$\vec{j}_c=-e\alpha_R\frac{\hat{z}\times\vec{S}}{\hbar}$,\\ $\vec{S}=\alpha_Rm^*\frac{\hat{z}\times\vec{j}_c}{\hbar}$} & Spin detection, spin-to-charge conversion \\ \hline
  Hall effect\cite{Liu2018} & Hall coefficient $R_H$ & $R_H=\frac{1}{pe}(1+\frac{64\pi {m^*}^2\gamma_R^2}{\hbar^4}p)$ & Spin dynamics observation \\
  \hline\hline
\end{tabular}
\label{table1}
\end{table*}
\vspace*{2mm}

However, the $\bf{k}$-linear Rashba effect for holes is not the reprint of electrons. Although their Hamiltonian has the same form, their effective spins differ due to their different orbital angular momentums. One example is the intrinsic spin Hall effect, which converts the unpolarized charge current to chargeless pure spin current (Table~\ref{table1})\cite{Sinova2004, DiXiao2010, Bernevig2005}. The spin Hall conductivity $\sigma_H$ is related to the Berry phase $\phi_B$, which is the surface integral of the Berry curvature $\Omega_p$ and the curvature is proportional to the $\bf{k}$-linear Rashba parameter and related to the spin and momentum. Compared with electrons, the Hamiltonian of holes has the same form, but the effective spin's difference will give rise to inequable Berry curvature, Berry phase, and spin Hall conductivity. Another example is the spin galvanic effect, where non-equilibrium spin density created by optical or electrical means is converted to a charge current (Table~\ref{table1})\cite{J.Wunderlich2009,Ivchenko1978,Ganichev2008}. Here, both the induced charge current density $\vec{j_c}$ and the non-equilibrium spin density $\vec{S}$ are related to the Rashba parameter, and the effective spin may make contributions as well. Besides, the Hall effect, where carriers accumulate perpendicular to both the electric field and magnetic field directions, is found to be modified by the $\bf{k}$-cubic Rashba effect in 2D systems (Table~\ref{table1})\cite{Liu2018}. This is because the effective magnetic field produced by spin-orbit coupling interacts with the external magnetic field. And our discovery of the $\bf{k}$-linear Rashba effect dominated over the $\bf{k}$-cubic effect will change the form of this effective magnetic field, resulting in the possibility of other forms of modification.

We hereby put forward a few examples of substantial effects above. More importantly, our newly-discovered $\bf{k}$-linear Rashba effect has distinct spin properties so that one has to reunderstand the relevant 2D hole spin physics based on the conventional $\bf{k}$-cubic Rashba model. Our discovery will force us to review all these substantial effects related to hole spins in 2D systems, further driving forward the development of devices. Furthermore, it remains an open question to detect this $\bf{k}$-linear spin splitting term experimentally.

\noindent {\bf \large Discussion}

Even in the absence of the direct dipolar coupling, Winkler\cite{Winkler2003} argued that the HH-LH mixing at $\bar{\bf{k}}=0$ will also produce a conventional $\bf{k}$-linear Rashba term to HH-like subbands. However, this conventional $\bf{k}$-linear Rashba SOC effect is small when compared to the $\bf{k}$-cubic term. Kloeffel\cite{Kloeffel2011} demonstrated that, in 1D nanowires, this conventional $\bf{k}$-linear Rashba SOC is in the third-order of multiband perturbation theory and hence different from the first-order direct Rashba SOC in both field and size-dependence. Specifically, the conventional $\bf{k}$-linear Rashba term is 10-100 times weaker than the direct Rashba SOC, and is stronger in narrower QWs, which is opposite to the results shown in Figure \ref{fig3}. We note that in a recent experiment the $\bf{k}$-linear Rashba SOC of the 2DHG is claimed to be absent in [001]-oriented strained Ge/SiGe QWs, where the weak antilocalization (WAL) feature in the magnetoconductivity measurement failed to be described by the $\bf{k}$-linear term alone but described well by the $\bf{k}$-cubic term alone\cite{Moriya2014}. Here, we have to stress that both $\bf{k}$-linear and $\bf{k}$-cubic terms are presented in our atomistic SEPM results, although the $\bf{k}$-linear term dominant over the $\bf{k}$-cubic term in an extreme small $\bar{k}$-range. Therefore, in fitting to WAL data, one has to include both $\bf{k}$-linear and $\bf{k}$-cubic terms instead of exclusively considering $\bf{k}$-linear or $\bf{k}$-cubic term only\cite{Moriya2014}.

In conclusion, we uncover a strong electric-tunable $\bf{k}$-linear Rashba SOC of 2DHGs in Ge/Si QWs. We illustrate that this previous unknown $\bf{k}$-linear Rashba SOC is a first-order direct Rashba effect, originating from a combination of HH-LH mixing and direct dipolar intersubband coupling. Specifically, in [110]-oriented Ge/Si QWs, the strength of this $\bf{k}$-linear Rashba SOC can be significantly enhanced by applied electric field to exceed 120 meV{\AA}, comparable to the largest values of 2D electron gases reported in narrow bandgap III-V semiconductors, facilitating the fast manipulation of hole spins.  This finding renders 2DHGs in Ge/Si QWs as an excellent platform for quantum computation. We have to stress that these findings are also applicable to 2DHGs in other tetrahedral semiconductors (see supplemental material). Our discovery makes a call to revisit the understanding of 2D hole spin physics, which have been explored with the assumption of $\bf{k}$-cubic Rashba SOC.

\vspace*{5mm}
\noindent {\bf \large Methods}

\textbf{Direct evaluation of the Rashba parameter based on the semi-empirical pseudopotential method.} Electronic structures are calculated using our previously developed atomistic, material-dependent semi-pseudopotential method (SEPM). An energy cutoff of 8.2 Ry, which is designed to fit the pseudopotential, is used to select the plane-wave basis, and fast Fourier transformations are used to transform the wave function between a real space grid and a reciprocal space grid. A $16\times16\times16$ grid in real space is used for each eight-atom cubic (diamond or zinc-blende) cell\cite{Wang1995, Wang1999}. The crystal potential consists of a superposition of screened atomic potentials, which contain a local part and a non-local spin-orbit interaction part. The atomic potentials are fit to experimental transition energies, effective masses, spin-orbit splitting, and deformation potentials of the bulk semiconductors to remove the ``LDA error"\cite{Wang1995, Wang1999}. The spin splitting energy is extracted from the calculated band structure, and the $\bf{k}$-linear and $\bf{k}$-cubic Rashba parameters are obtained by fitting to the spin splitting.

\vspace*{5mm}
\noindent {\bf \large Data availability.}

The data that support the findings of this study are available from the corresponding author upon reasonable request.

\vspace*{10mm}
\noindent {\bf \large References}

%\bibliography{Discovery_ref}

\vspace*{4mm}
\noindent {\bf \large Acknowledgements}

\noindent The authors thank Professor X. W. Zhang for helpful discussion. The work was supported by the National Science Fund for Distinguished Young Scholars under grant No. 11925407, the Basic Science Center Program of the National Natural Science Foundation of China (NSFC) under grant No. 61888102, and the Key Research Program of Frontier Sciences, CAS under grant No. ZDBS-LY-JSC019. S. G. was also supported by the NSFC under grant No. 11904359.

\vspace*{5mm}
\noindent {\bf \large Author contributions}

\noindent J. X. performed the electronic structure calculations, prepared the figures, and developed the effective models with the help of S. G. J. L. proposed the research project. J. L. and S. L. established the project direction and conducted the analysis, discussion, and writing of the paper with input from J. X.

\vspace*{5mm}
\noindent {\bf \large Competing interests:}
%\noindent {\bf \large Author contribution}

\noindent The authors declare no competing interests.

\vspace*{5mm}
\noindent {\bf \large Additional information}

\noindent {\bf Supplementary information} is available in the online version of the paper. Reprints and permissions information is available online at www.nature.com/reprints. Correspondence and requests for materials should be addressed to S. G. (Email: shan\_guan@semi.ac.cn) and J.-W. L. (Email: jwluo@semi.ac.cn).

\end{document}